\documentclass[aps,prl,reprint,superscriptaddress,longbibliography]{revtex4-1}

\usepackage{graphicx}
\usepackage{amsmath}
\usepackage{amssymb}
\usepackage{nicefrac}

\usepackage{mathtools}
\usepackage[per-mode=symbol, separate-uncertainty = true, multi-part-units = single]{siunitx}
\usepackage{xspace}		
\usepackage[hidelinks]{hyperref} 

\usepackage{booktabs}

\usepackage[version=3]{mhchem}
\usepackage{siunitx}
\usepackage[normalem]{ulem}
\usepackage{textcomp}
\usepackage{upgreek}

\let\orgautoref\autoref
\providecommand{\Autoref}{%
  \def\sectionautorefname{Section}%
  \def\figureautorefname{Figure}%
  \def\subfigureautorefname{Figure}%
  \orgautoref}
\renewcommand{\autoref}{%
  \def\sectionautorefname{Sec.}%
  \def\figureautorefname{Fig.}%
  \def\subfigureautorefname{Fig.}%
  \orgautoref}

\usepackage{color}
\usepackage[normalem]{ulem}
\definecolor{darkgreen}{rgb}{0.0,0.7,0.0}
\newcommand{\rtext}[1]{\textcolor{red}{{#1}}}

\begin{document}


\title{Hyperfine-structure-induced depolarization of impulsively aligned  ${\rm {\bf I_{2}}}$ molecules}



\author{Esben F. Thomas}
\affiliation{Department of Chemistry, Technical University of Denmark, Building 206, DK-2800 Kongens Lyngby, Denmark}

\author{Anders A. S{\o}ndergaard}
\affiliation{Department of Chemistry, Aarhus University, Langelandsgade 140, DK-8000 Aarhus C, Denmark}

\author{Benjamin Shepperson}
\affiliation{Department of Chemistry, Aarhus University, Langelandsgade 140, DK-8000 Aarhus C, Denmark}

\author{Niels E. Henriksen}
\affiliation{Department of Chemistry, Technical University of Denmark, Building 206, DK-2800 Kongens Lyngby, Denmark}

\author{Henrik Stapelfeldt}
\affiliation{Department of Chemistry, Aarhus University, Langelandsgade 140, DK-8000 Aarhus C, Denmark}




\begin{abstract}
A moderately intense $450$ fs laser pulse is used to create rotational wave packets in gas phase $\rm{I_2}$  molecules. The ensuing time-dependent alignment, measured by Coulomb explosion imaging with a delayed probe pulse, exhibits the characteristic revival structures expected for rotational wave packets but also a complex non-periodic substructure and decreasing mean alignment not observed before. A quantum mechanical model attributes the phenomena to coupling between the rotational angular momenta and the nuclear spins through the electric quadrupole interaction. The calculated alignment trace agrees very well with the experimental results.\end{abstract}



\maketitle

Alignment of isolated molecules, i.e. confinement of their internal axes to directions fixed in space, by moderately intense laser pulses is considered a well-understood process resulting from the polarizability interaction~\cite{stapelfeldt_colloquium:_2003,seideman_nonadiabatic_2005,ohshima_coherent_2010,fleischer_molecular_2012}. In the impulsive limit, where a laser pulse much shorter than the molecular rotational period is used, each molecule is left in a superposition of rotational eigenstates. For the widely studied case of linear molecules and a linearly polarized fs alignment pulse, this wave packet formation causes the molecules to align 
shortly after the laser pulse and in periodically occurring narrow time windows, termed revivals~\cite{seideman_revival_1999,rosca-pruna_experimental_2001,machholm_field-free_2001,renard_postpulse_2003,dooley_direct_2003}. In the rigid rotor approximation the revival pattern repeats itself~\cite{przystawik_generation_2012} unless the rotational coherence is distorted by e.g. a dissipative environment~\cite{ramakrishna_intense_2005,vieillard_field-free_2008,owschimikow_cross_2010,hartmann_quantum_2012,pentlehner_impulsive_2013,shepperson_laser-induced_2017}.


Decades of frequency-resolved high-resolution spectroscopy~\cite{gordy_microwave_1984,Zare1988} and time-dependent depolarization experiments on molecules prepared in single rotational states (see Refs.~\citep{Code71,Fano73,altkorn_depolarization_1985,Yan93,Gough93,Cool95,Zhang96,Wouters97,Rudert99,Sofikitis07,Bartlett09,Bartlett10,Grygoryeva17} for previous examples) have, however, established that a rigid rotor model is insufficient and that a precise description of rotational spectra must include the coupling between rotational angular momentum and electronic or nuclear spin. It is, therefore, surprising that the influence of such effects, notably the hyperfine coupling between the electric quadrupole moment of the nuclei and the electric field of the electrons, has never been addressed in fs-laser-induced molecular alignment studies. In the current work we measured the time-dependent degree of alignment, induced by a $450$ fs pulse, for a sample of $\rm{I_2}$ molecules covering the first seven rotational revivals. By contrast to the aforementioned depolarization studies, which do not involve coherent superpositions of rotational states, our experiment probes the impact of hyperfine coupling on the revival structures.


Using a quantum mechanical model in conjunction with the experimental results, we find that the hyperfine coupling affects the revival structures in qualitatively different ways compared to the well-understood impact on the ``permanent'' alignment of a molecule prepared in a single rotational state. Notably, the effect on the permanent alignment is known to be negligible in the limit where the rotational angular momentum is much larger than the angular momentum of the total nuclear spin~\citep{Bartlett09}. By contrast, we find that the hyperfine coupling will always significantly perturb the revival structures over time.

The experimental setup and methods were described previously~\cite{shepperson_strongly_2017}, so only a few details are pointed out here. A pulsed molecular beam, formed by expanding ~$\sim 1$~mbar iodine gas in $80$ bar of He gas into vacuum, enters a velocity map imaging (VMI) spectrometer where it is crossed at $90^{\circ}$ by two pulsed collinear laser beams. The first pulse (kick pulse, $\lambda = 800$ nm, $\tau_{\rm FWHM} = 450$ fs, I$_\text{0}$ = $\SI{1.1e12}{W/cm^2}$) creates rotational wave packets in the $\rm{I_2}$ molecules. The second pulse (probe pulse, $800$ nm, $35$ fs, $\SI{4.3e14}{W/cm^2}$) Coulomb explodes the molecules. This leads to $\rm{I^+}$ ion fragments with recoil directions given by the angular distribution of the molecular axes at the instant of the probe pulse. The emission directions of the $\rm{I^+}$ ions are recorded with a $2$D imaging detector at different kick-probe delays, which allows us to determine the time-dependent degree of alignment, $\langle \cos^2 \theta_{\rm 2D} \rangle$,
$\theta_{\rm 2D}$ being the angle between the alignment pulse polarization and the projection of an $\rm{I^+}$ ion velocity vector on the detector~\citep{Soendergaard17}.

\begin{figure*}
\noindent \centering{}\includegraphics[scale=0.455]{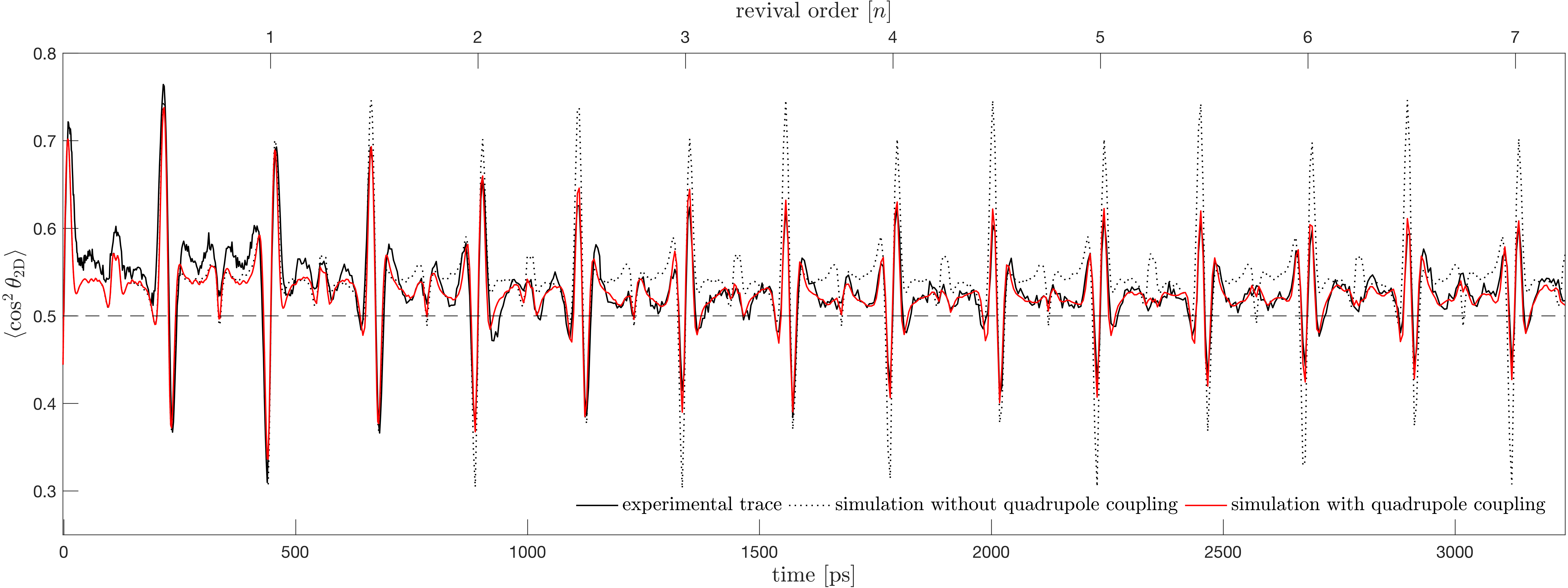}\caption{Experimental results (black) and the model calculations
(red) calculated at $T=0.8$ K. An alignment
trace calculated without quadrupole coupling is also shown (dotted black).\label{Trace_Plot}}
\end{figure*}

The time dependence of $\langle \cos^2 \theta_{\rm 2D} \rangle$ determined experimentally is shown in black in \autoref{Trace_Plot}. The alignment trace is dominated by the pronounced half and full revivals, but their amplitude decreases with the revival order and their structure is changing. These observations are not caused by experimental factors such as collisions or limited temporal detection windows (see Supplemental Material~\cite{supplement}, which includes Refs.~\cite{NIST,Bacis1980,Zare1982,Maroulis1992,Maroulis1997,even_cooling_2000,filsinger_quantum-state_2009,Shu2017}). For comparison, $\langle \cos^2 \theta_{\rm 2D} \rangle$ calculated by solving the time-dependent Schr\"{o}dinger equation (TDSE) for a rigid rotor is also shown. It is clear that the decreasing amplitude and changing structure of the revivals observed experimentally is at odds with the calculations. In particular, significant experimental deviations from the calculated results are evident in the higher order fractional revivals, for instance at the 6 + $\nicefrac{3}{4}$ revival the experimental and calculated peaks point in opposite directions.

We now show that the numerical results match the experimental findings to a high degree of accuracy when hyperfine coupling is included in the theoretical model. For $\rm{I_2}$ molecules this mainly stems from the coupling between the electric quadrupole moment of the atomic nuclei and the gradient of the electric field created by the electrons. The coupling between the magnetic dipole moment of the nuclei and the B-field from the electrons is much weaker and not included in our model~\cite{Yokozeki80}. The total nuclear spin, ${\bf I}$, is the sum of the spins of the two atomic nuclei: ${\bf I}$ = ${\bf I}_{1} + {\bf I}_{2}$. As a result there are $36$ nuclear spin isomers $\left|I M_{I}\right\rangle$ with $0\leq I\leq5$ and $-I\leq M_{I}\leq I$ since the nuclear spin of $^{127}{\rm I}$ is $5/2$. We assume that the $36$ nuclear spin isomers are initially equally abundant \citep{McQuarrie1976}. The rotational wave packet created by the alignment pulse from an initial rotational eigenstate $\left|J_i M_i \right\rangle$ is denoted $\sum_{J}a_{J}\left|JM_{i}\right\rangle$ ($M_i$ is not changed due to the linear polarization of the alignment pulse). The symmetry requirements of the total molecular wave function entail
that the parity of the $I$ and $J$ states must be the same in a given molecule \citep{McQuarrie1976}. Consequently, there are $15$/$21$ para/ortho
(even/odd $I$ and $J$) spin isomers. The coupled spin isomer -- rotational wave packet is described as:
\begin{multline}
\left|IM_{I}\right\rangle \otimes\sum_{J}a_{J}\left|JM_{i}\right\rangle =\\
\sum_{J} a_{J} \sum_{F}C_{M_{I}M_{i}M_{F}}^{IJF}\left|IJFM_{F}\right\rangle ,\label{fully_coupled}
\end{multline}
where $F$ is the total angular momentum, ${\bf F}$ = ${\bf I} + {\bf J}$, $M_{F}=M_{I}+M_{i}$,
and $C_{M_{I}M_{i}M_{F}}^{IJF}$ the Clebsch-Gordan coefficients.

In preparation for solving the TDSE, we construct a square matrix ${\bf H}$ in the $\left|IJFM_{F}\right\rangle$ basis, with elements given by \citep{Cook1971}:
\begin{multline}
H_{a,b}=\left\langle I^{a}J^{a}F^{a}M_{F}^{a}\left|\mathcal{H}_{B}+\mathcal{H}_{Q}\right|I^{b}J^{b}F^{b}M_{F}^{b}\right\rangle \\
=\delta_{J^{b}J^{a}}B_0J^{a}\left(J^{a}+1\right)\\
-\left(eqQ\right)\delta_{F^{b}F^{a}}\delta_{M_{F}^{b}M_{F}^{a}}\left[\left(-1\right)^{I^{a}}+\left(-1\right)^{I^{b}}\right]\left(-1\right)^{F^{a}+I^{a}}\\
\times\left[\frac{21}{20}\left(2I^{b}+1\right)\left(2I^{a}+1\right)\left(2J^{b}+1\right)\left(2J^{a}+1\right)\right]^{1/2}\\
\times\begin{pmatrix}J^{b} & 2 & J^{a}\\
0 & 0 & 0
\end{pmatrix}\begin{Bmatrix}F^{a} & I^{a} & J^{a}\\
2 & J^{b} & I^{b}
\end{Bmatrix}\begin{Bmatrix}\frac{5}{2} & I^{a} & \frac{5}{2}\\
I^{b} & \frac{5}{2} & 2
\end{Bmatrix},\label{matrix_element}
\end{multline}
 $\mathcal{H}_{B}$ describes the rigid rotor Hamiltonian, where $B_0=1.11863$ GHz \cite{NIST} is the molecular rotational constant of $\rm I_2$ in the vibrational ground state (centrifugal distortion was found to be negligible~\cite{supplement}), and $\mathcal{H}_{Q}$ is the electric quadrupole interaction component of the hyperfine structure Hamiltonian for a
diatomic molecule, where $eqQ=-2.45258$ GHz \cite{Yokozeki80} is the quadrupole coupling constant.

$\mathcal{H}_{Q}$ introduces shifts in the diagonal elements of $\bf H$, and off-diagonal couplings when $\Delta I=\pm2$ and/or $\Delta J=\pm2$. Generally, $\bf H$ must incorporate all initial states occupied at $t=0$ (i.e. those given by the right side of \autoref{fully_coupled}), as well as states that may become occupied over time as a result of the off-diagonal couplings. Specifically, it was found that any states that can be reached via inter-$I$ coupling must be included to faithfully reproduce the experimental alignment trace; however, inter-$J$ couplings were found to have a negligible impact on $\langle \cos^2 \theta_{\rm 2D} \rangle$ (attributed to the relatively large energy differences between various $J$ states). As such, the $\left|IJFM_{F}\right\rangle$ states incorporated in $\bf H$ need only contain $J$-values that were already present in $\sum_{J}a_{J}\left|JM_{i}\right\rangle$.

The TDSE is solved by expanding the
full wave function $\Psi$ onto the coupled basis functions, i.e.
$\Psi\left(t\right)=\sum_{k=1}^{N}c_{k}\left(t\right)\left|I^{k}J^{k}F^{k}M_{F}^{k}\right\rangle $ where $N$ is the order of $\bf H$ and $\Psi\left(0\right)$ is given by \autoref{fully_coupled}.
The $c_{k}\left(t\right)$ coefficients are found by
diagonalizing ${\bf H}$ to solve the resulting system of coupled
linear differential equations. $\Psi\left(t\right)$ is
then transformed back into the uncoupled representation to calculate
the alignment trace, i.e.
\begin{multline}
\Psi\left(t\right)=\sum_{k=1}^{N}c_{k}\left(t\right)\sum_{M_I,M_J} C_{M_{F}^kM_{I}M_{J}}^{F^kI^kJ^k}\left|I^kM_{I}\right\rangle \otimes\left|J^kM_{J}\right\rangle, \label{Psi_t}
\end{multline}
Efficient calculation of $\langle\cos^{2}\theta_{{\rm 2D}}\rangle$
is achieved by expanding $\cos^{2}\theta_{{\rm 2D}}$ onto a basis
of Legendre polynomials as described in Ref. \citep{Soendergaard17}, and noting
the orthonormality of the $\left|IM_{I}\right\rangle $ states in
\autoref{Psi_t}.

The alignment trace of any initial $\sum_{J}a_{J}\left|JM_{i}\right\rangle$ superposition is the equally weighted, incoherent sum of traces generated by coupling to all nuclear spin isomers that symmetry requirements will permit. The \emph{complete} alignment trace is the weighted incoherent sum of traces from all the \emph{different} initial $\sum_{J}a_{J}\left|JM_{i}\right\rangle $ superpositions that exist because of thermal and focal volume averaging, as per the methodology outlined in Refs. \citep{BisgaardPHD,SoondergaardPHD}.

The simulated alignment trace with the effects of quadrupole coupling included is shown in red in \autoref{Trace_Plot}. Based on previous work in our group, we estimate that the molecules are initially in thermal (Boltzmann) equilibrium at $0.8$ K~\cite{shepperson_strongly_2017}. The minor discrepancy between the theoretical and experimental traces from $0-600$ ps can essentially be eliminated by fitting the temperature~\cite{supplement}, however this results in a fitted temperature of $\sim 0.4$ K, which we believe is unrealistically low.

Let $\sum_{J,J^{\prime}}\left\langle JM_{J}\left|\cos^{2}\theta_{{\rm 2D}}\right|J^{\prime}M_{J}\right\rangle $ represent the sum of matrix elements that generates the theoretical alignment trace, where we omit the nuclear spin states and time-dependent coefficients in \autoref{Psi_t} for clarity. The $J\neq J^{\prime}$ terms are sinusoidal functions oscillating at frequencies proportional to the energy difference between the $J$ and $J^{\prime}$ states. These terms represent the coherence of the wave packet and are responsible for the revivals. In analogy with previous work~\cite{ramakrishna_intense_2005} we refer to their sum as $\langle \cos^{2}\theta_{{\rm 2D}}\rangle_{\rm coh}$. Conversely, the terms where  $J= J^{\prime}$ represent the population of the rotational states and characterize the permanent alignment. Their sum is denoted $\langle \cos^{2}\theta_{{\rm 2D}}\rangle_{\rm perm}$~\cite{ramakrishna_intense_2005}. Note that the mean alignment of the trace is well characterized by $\langle \cos^{2}\theta_{{\rm 2D}}\rangle_{\rm perm}$, as this term provides the baseline value around which $\langle \cos^{2}\theta_{{\rm 2D}}\rangle_{\rm coh}$ oscillates.

\begin{figure*}
\noindent \centering{}\includegraphics[scale=0.455]{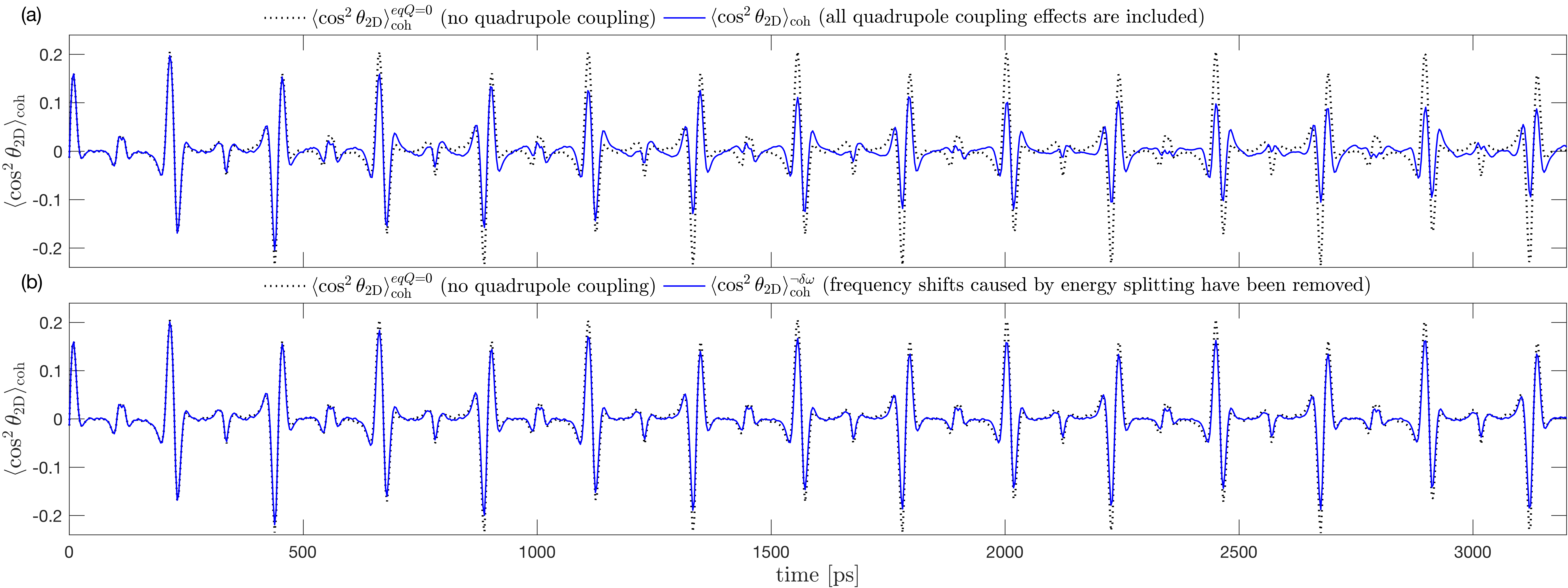}\caption{In (a) the sum of $J\neq J^{\prime}$ matrix elements comprising the theoretical quadrupole coupled alignment trace, $\langle \cos^{2}\theta_{{\rm 2D}}\rangle_{\rm coh}$, is shown in blue. The blue trace in (b) shows $\langle \cos^{2}\theta_{{\rm 2D}}\rangle_{\rm coh}^{\neg \delta \omega}$, where the frequency and phase shifts caused by the hyperfine energy splitting have been suppressed. Both (a) and (b) are superposed with equivalent traces calculated without quadrupole coupling (dotted black).  \label{AC_DC_traces}}
\end{figure*}

Visual inspection of \autoref{Trace_Plot} indicates that the mean alignment is slightly decreasing from $0-1000$ ps. This behaviour is attributed to the well-understood fact that quadrupole coupling leads to changes in the $M_{J}$ projection of a single $J$ state due to angular ``precession'' of the coupled $\bf I$ and $\bf J$ vectors around $\bf F$  (see, e.g., Fig. $1$ in Ref. \citep{Bartlett10}) and changes in the relative orientation of the individual nuclear spin vectors (which we denote ``spin flipping''). Previous experiments on hyperfine-induced depolarization of single rotational states have shown that this effect (hereafter referred to as ``precession-type depolarization'') leads to a general time-dependent decrease in molecular alignment \citep{Code71,Fano73,Yan93,Gough93,Cool95,Zhang96,Wouters97,Rudert99,Sofikitis07,Bartlett09,Bartlett10,Grygoryeva17}. \Autoref{AC_DC_traces}(a) shows $\langle \cos^{2}\theta_{{\rm 2D}}\rangle_{\rm coh}$ superposed with the sum of $J\neq J^{\prime}$ trace elements calculated without quadrupole coupling (denoted $\langle \cos^{2}\theta_{{\rm 2D}}\rangle_{\rm coh}^{eqQ = 0}$) for comparison. It is seen that the quadrupole coupling also strongly affects the revival structures.

To understand the cause of the amplitude loss and substructure modification in  $\langle \cos^{2}\theta_{{\rm 2D}}\rangle_{\rm coh}$, the effects of precession-type depolarization were artificially suppressed in the model by changing all $M_{J}$'s in \autoref{Psi_t} to the $M_{i}$ from the initial $\sum_{J}a_{J}\left|JM_{i}\right\rangle$ superposition when calculating the $\left\langle JM_J  \left|\cos^{2}\theta_{{\rm 2D}}\right| J^{\prime}M_J \right\rangle$ elements (while treating everything else as if the ``actual''  $M_{J}$'s are still in place). Surprisingly, this has very little effect on the shape of $\langle \cos^{2}\theta_{{\rm 2D}}\rangle_{\rm coh}$, indicating that some previously unexplored mechanisms associated with the quadrupole coupling are causing the modulations in the signal.

\begin{figure*}
\noindent \centering{}\includegraphics[scale=0.455]{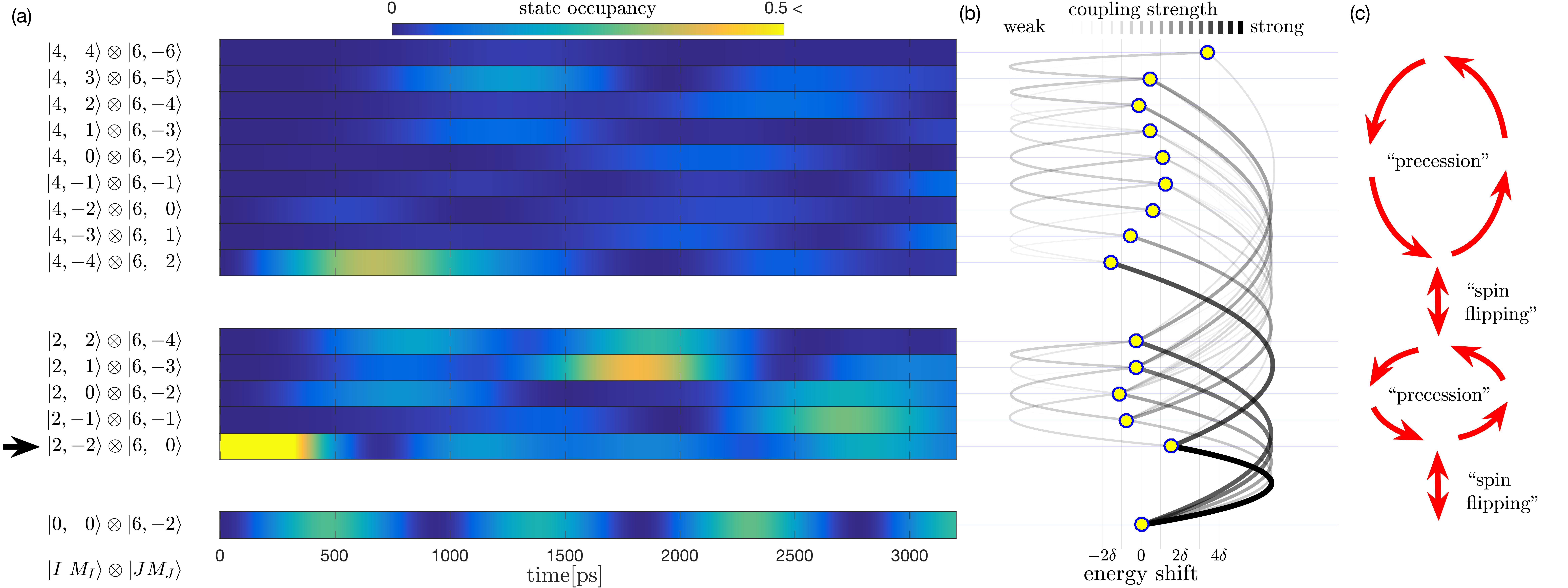}\caption{(a) Occupancy of initial state $\left|2,-2\right\rangle \otimes\left|6,0\right\rangle $
and the $\left|IM_{I}\right\rangle \otimes\left|JM_{J}\right\rangle$ states it couples to evolving over the timescale of the experiment.
(b) Sketch of the state energy splittings and relative coupling strengths ($\delta = 57$ MHz). (c) Schematic classical interpretation of the system dynamics.
 \label{State_Dyn}}
\end{figure*}

It is informative to show the quadrupole coupled dynamics of a molecule starting in a single spin isomer/$J$ state combination. In \autoref{State_Dyn}(a) the time evolution of initial state $\left|2,-2\right\rangle \otimes\left|6,0\right\rangle$ is shown projected onto the $\left|IM_{I}\right\rangle \otimes\left|JM_{J}\right\rangle$ basis, where the (negligible) effect of inter-$J$ coupling has been suppressed for clarity. The example in \autoref{State_Dyn} illustrates how the quadrupole coupling will cause each $\left|JM_{i}\right\rangle$ from the initial $\sum_{J}a_{J}\left|JM_{i}\right\rangle$ superposition to spread out across a ``$J$-manifold'' of coupled states. Also, \autoref{State_Dyn} shows how all states in a given $J$-manifold will have different $\left|IM_{I}\right\rangle$. Therefore, orthonormality of the $\left|IM_{I}\right\rangle$ spin states implies that each state in the $J$-manifold will combine with at \emph{most} one state in the $J^{\prime}$-manifold to yield a nonzero contribution to the alignment trace.

Given two or more superposed $J$ states coupled to the same spin isomer at $t=0$,  the energy splitting, coupling strength, and number of states associated with each $J$-manifold partially depends on $J$ (attributable, e.g., to the appearance of $J^{a,b}$ in \autoref{matrix_element}). Dissimilarities in the energy splitting between different $J$-manifolds introduces multiple frequency shifts into the components of $\langle \cos^{2}\theta_{{\rm 2D}}\rangle_{\rm coh}$. The beating caused by the introduction of these new frequencies modulates the alignment trace. We investigated the nature of this frequency beating by artificially suppressing its effect in the model. This was done by eliminating the quadrupole-coupling-induced frequency and phase shifts in the complex arguments of the coefficients governing the time evolution of all $\left|IM_{I}\right\rangle \otimes\left|JM_{J}\right\rangle$ states across all $J$-manifolds. In this way we calculate a modified trace, $\langle \cos^{2}\theta_{{\rm 2D}}\rangle_{\rm coh}^{\neg \delta \omega}$, where $\neg \delta \omega$ indicates that all frequency shifts introduced by the energy splitting have been removed while leaving the $J$-manifold population dynamics unchanged. A plot of $\langle \cos^{2}\theta_{{\rm 2D}}\rangle_{\rm coh}^{\neg \delta \omega}$ is shown in \autoref{AC_DC_traces}(b). Comparison of the $\langle \cos^{2}\theta_{{\rm 2D}}\rangle_{\rm coh}$ and $\langle \cos^{2}\theta_{{\rm 2D}}\rangle_{\rm coh}^{\neg \delta \omega}$ traces shown in \autoref{AC_DC_traces} reveals that the frequency beating plays a significant, but not singular, role in attenuating the peak amplitudes. It is also remarkable that the higher order fractional revivals in $\langle \cos^{2}\theta_{{\rm 2D}}\rangle_{\rm coh}^{\neg \delta \omega}$ do not exhibit the deviations and sign changes that are present in  $\langle \cos^{2}\theta_{{\rm 2D}}\rangle_{\rm coh}$. This demonstrates that the complex non-periodic substructures observed in the experimental trace can be solely attributed to the new frequencies introduced into $\langle \cos^{2}\theta_{{\rm 2D}}\rangle_{\rm coh}$ by the hyperfine coupling.

Note that the peak amplitudes in $\langle \cos^{2}\theta_{{\rm 2D}}\rangle_{\rm coh}^{\neg \delta \omega}$ still decrease compared to $\langle \cos^{2}\theta_{{\rm 2D}}\rangle_{\rm coh}^{eqQ = 0}$. This is because, as stated earlier, the state population distributions of manifolds with different $J$ will become increasingly dissimilar over time. These asynchronous distributional dynamics cause a net loss of amplitude due to the bijective/injective (one to at most one) way of combining different sets of states associated with different $J$-manifolds when calculating nonzero contributions to the trace. Experimentally, some of the observed peak attenuation may in principle be caused by $\rm I_2$ molecules in vibrationally excited states. Our analysis shows, however, that the potential impact is minor~\cite{supplement}.

It has been remarked that precession-type depolarization in single $J$ states is most significant when the coupled $\bf I$ and $\bf J$ vectors have similar magnitudes~\citep{Bartlett09}. Conversely, our analysis suggests that the observed modulations in the revival structures of $\langle \cos^{2}\theta_{{\rm 2D}}\rangle_{\rm coh}$ are not directly contingent on the magnitude of  $\bf I$ or $\bf J$. Therefore, we investigate what happens if rotational wave packets containing larger $J$ are created. To this end, we simulated the effects of quadrupole coupling in $\rm I_2$ molecules aligned with pulses up to $9\times$ more intense than used in the current experiment.

\begin{figure}
\noindent \centering{}\includegraphics[scale=0.44]{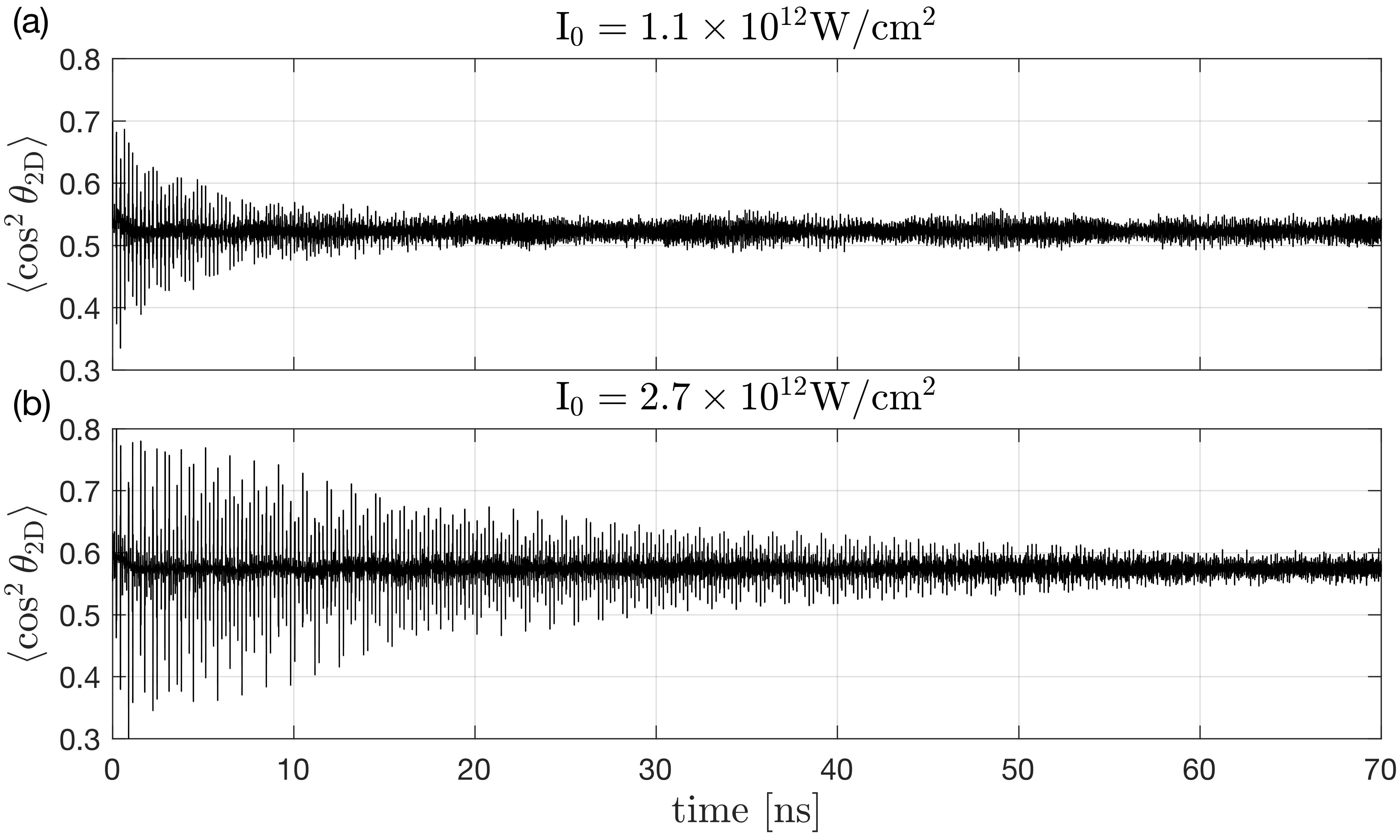}\caption{Quadrupole-coupled alignment traces simulated to $70$ ns, with alignment pulse intensities set to (a) the experimental and (b) $2.5\times$ the experimental value.
 \label{decay}}
\end{figure}

Increasing the pulse intensity leads to initial revival peaks with larger amplitudes, as well as a higher level of mean alignment. The early decrease in mean alignment observed in the experiment becomes less pronounced at higher intensities, and for all intensities the mean alignment is nearly constant when $t>1$ ns. Additionally, it was found that for all intensities the revival structures \emph{always} decay into what resembles low amplitude unstructured ``noise'', however this decay takes longer for more intense pulses, as illustrated in \autoref{decay}~\cite{supplement}. These observations agree qualitatively with our expectations, i.e. the classical model of precession predicts that $\langle \cos^{2}\theta_{{\rm 2D}}\rangle_{\rm perm}$ will change less for wave packets with large $J$, whereas the scrambling/attenuating effects of the frequency beating and asynchronous dynamics will accumulate over time and eventually dominate the $\langle \cos^{2}\theta_{{\rm 2D}}\rangle_{\rm coh}$ component of the trace regardless of the magnitude of the $J$-values present in the wave packet. \rtext{}

In closing, we note that the alignment trace of any molecule containing heavy atoms (e.g. $\rm I$ or $\rm Br$) with large quadrupole coupling constants is expected to show similar deviations from the rigid rotor approximation when excited into a coherent superposition of rotational eigenstates.


HS acknowledges support from the European Research Council-AdG (Project No. 320459, DropletControl).



%

\end{document}